\documentclass[10pt,conference]{IEEEtran}
\usepackage{epsfig,rotating,setspace,latexsym,amsmath,epsf,amssymb,amsfonts,bm,theorem,cite,authblk, bbm,color, multirow, caption, subcaption, booktabs, xcolor}
\IEEEoverridecommandlockouts
\allowdisplaybreaks

\title{Spectrum Adaptation of AI-Native Autoencoder Communications with Transfer Learning} 

\author{Yalin E. Sagduyu and Tugba Erpek}
\affil{\normalsize Nexcepta, Gaithersburg, MD, USA}

\date{}
\IEEEoverridecommandlockouts

\begin{document}
	
\maketitle
\thispagestyle{empty}
	
\begin{abstract}
End-to-end physical-layer learning replaces separate modulation, coding, and detection operations with an autoencoder (AE) that consists of a jointly trained encoder-decoder pair at the transmitter and the receiver. This paper studies AI-native communications with AEs and quantifies the sensitivity to training-test mismatch in channel models. Models trained only on an AWGN channel achieve low bit error rates (BERs) under matched conditions but degrade severely on Rayleigh fading with practical front-end impairments. To reduce the time and energy required to adapt to new channel conditions, we propose a low-latency transfer-learning procedure that starts from an AWGN-pretrained system and fine-tunes only a small subset of transmitter and receiver layers while freezing the remainder. We further formulate layer selection as a budget-constrained optimization problem that identifies the most informative parameters to update during adaptation. This preserves the learned signal geometry and the deep decision structure while adapting lightweight front ends to fading- and impairment-induced mixing. For AE communications, we show that partial updates remove the mismatch-induced error floor and approach the BER of fully retrained models. We provide a layer-wise study and quantify resource-performance trade-offs, showing that modest adaptation substantially reduces BER while cutting training time, multiply-accumulate (MAC) operations, energy consumption, and the fraction of updated parameters for resource-efficient implementation.
\end{abstract}
	
	
\section{Introduction}
End-to-end learned physical-layer transmitters and receivers are the key building blocks of \emph{AI-native communications}, enabling joint optimization of waveform generation and detection by modeling the entire link as a neural \emph{autoencoder} (AE) trained over a wireless channel \cite{OShea2017TCCN, erpek2019deep}. For short blocklengths and practical signal-to-noise ratio (SNR) ranges, these designs achieve bit-error rates (BERs) comparable to, and in some regimes better than, conventional modulation and coding with handcrafted receivers \cite{AitAoudia2022_TWC_OFDM,Honkala2021_TWC_DeepRx,Song2022_TWC_Benchmarking}. In doing so, learned transceivers naturally discover irregular constellations, shaping gains, and task-specific invariances without committing to a fixed code or constellation structure.

A key practical limitation is that the learned transmitter-receiver pair is implicitly matched to the training channel. When the deployment channel deviates from this distribution (e.g., an AWGN-trained system operating over Rayleigh fading with impairments), the receiver decision regions and transmitter signaling geometry become misaligned, leading to elevated BER or error floors unless the model is adapted. Despite advances in data-driven channel learning, learned transceivers remain sensitive to training-test channel mismatch \cite{Ye2020_TWC_cGANChannel}.

A naive remedy is to fully retrain the model on the new channel, but this is time- and energy-intensive. From random initialization, the optimizer must relearn signaling geometry and equalization/decision behavior, especially in multi-antenna fading environments. In contrast, \emph{transfer learning} amortizes pretraining cost and updates only what is necessary for the target environment. Prior physical-layer learning studies show that knowledge transfer reduces data and computation for tasks such as channel state prediction and compression/feedback \cite{Yang2020_TCOM_TransferDL,liu2024deep,Li2024_TWC_KnowledgeTransfer}, motivating similar strategies for end-to-end transceivers.
Such low-latency adaptation is a key enabler for AI-native mission-critical communications, allowing learned transceivers to rapidly adjust to changing operational conditions in the spectrum while maintaining reliable connectivity.

\begin{figure}[t!]
	\centering
	\includegraphics[width=1\columnwidth]{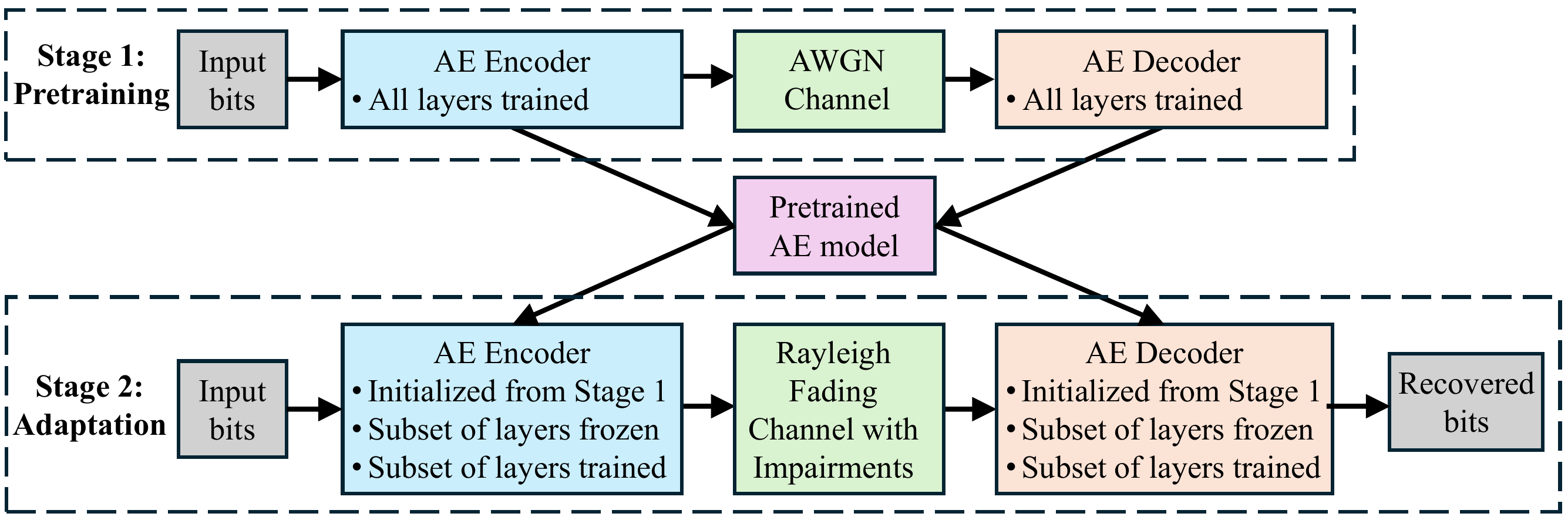}
	\caption{System model for AE communications with transfer learning.}
	\label{fig:AE_TL_system}
\end{figure}

In this paper, we consider low-delay adaptation of an AI-native AE-based transmitter-receiver from an AWGN source domain to a MIMO Rayleigh fading target domain, as shown in Fig.~\ref{fig:AE_TL_system}. The system is first trained on an AWGN channel across a broad SNR range to learn power-normalized signaling robust to additive noise. We then apply \emph{partial} fine-tuning in the Rayleigh channel domain by updating only a small, layer-indexed subset of parameters near the inputs of the transmitter and receiver, freezing the deeper layers. This design seeks to preserve the AWGN-learned signal geometry and deep decision structure while letting a lightweight front end become fading and impairment-aware.

Prior work has explored (i) the process of pretrain-then-fine-tune over the real channel for AE systems \cite{Cammerer2020_TCOM_Trainable}, (ii) model-free on-channel adaptation without an explicit channel model \cite{AitAoudia2019_JSAC_ModelFree}, and (iii) surrogate-channel pretraining using conditional generative adversarial networks (GANs) followed by fine-tuning under unknown or time-varying channels \cite{Ye2020_TWC_cGANChannel}. In contrast, this paper makes the following contributions by adapting from an AWGN channel to Rayleigh fading with impairments:  
(i) We formulate channel adaptation as a constrained layer-wise optimization and propose a low-delay transfer-learning scheme that updates only a subset of transmitter and receiver layers while freezing the remainder. 
(ii) We present results on a \(4{\times}4\) Rayleigh MIMO channel with impairments, quantifying BER and adaptation cost in terms of time, multiply-accumulate (MAC) operations, energy, and trainable-parameter fraction, and showing that limited partial updates sustain low BER at a fraction of the cost of full retraining.  
(iii) We introduce a budget-constrained layer-selection procedure that identifies the most informative parameters to update under a global retraining budget and improves BER compared with prefix-based layer updates under comparable parameter budgets.

The remainder of the paper is organized as follows. Sec.~\ref{sec:system_model} describes the AE communication model. Sec.~\ref{sec:mismatch} analyzes training-test mismatch. Sec.~\ref{sec:lowlatency} presents the proposed low-latency transfer-learning method. Sec.~\ref{sec:conclusion} concludes the paper.

\section{SISO and MIMO Autoencoder Communications} \label{sec:system_model}
We model the transmitter and receiver as an encoder-decoder pair communicating over a wireless channel. A binary vector \(\mathbf{b}\in\{0,1\}^k\) is mapped by the transmitter to a length-\(n\) complex waveform across \(N_t\) antennas, and the receiver observes a noisy superposition on \(N_r\) antennas to produce bitwise decisions.  The transmitter implements an encoder mapping
\begin{equation}
f_{\boldsymbol{\phi}}:\{0,1\}^k \rightarrow \mathbb{C}^{N_t\times n}, \qquad
\mathbf{x} = f_{\boldsymbol{\phi}}(\mathbf{b}),
\end{equation}
with parameter vector \(\boldsymbol{\phi}\). In implementation, the complex codeword \(\mathbf{x}\) is represented as an I/Q tensor of size \(n\times N_t\times 2\). After generation, each codeword is rescaled such that its total energy is approximately fixed:
\begin{equation}
\|\mathbf{x}\|_F^2 = \sum_{t=1}^{n}\sum_{m=1}^{N_t} |x_{m,t}|^2 \approx n.
\end{equation}
This normalization enforces constant average energy per channel use, making the bit energy depend only on \((k,n)\) and independent of the number of transmit antennas since total energy is fixed across antennas. We set \(k=8\) and \(n=16\), yielding an effective rate \(R_b = k/n = 0.5\) bits per complex channel use. The channel acts on $\mathbf{x}$ as
\begin{equation}
\mathbf{y}_t = \mathbf{H}_t \mathbf{x}_t + \mathbf{w}_t,\qquad t=1,\dots,n, \label{eq:channel}
\end{equation}
where $\mathbf{H}_t\in\mathbb{C}^{N_r\times N_t}$ and $\mathbf{w}_t \sim \mathcal{CN}(\mathbf{0}, N_0\mathbf{I})$
(i.e., variance $N_0/2$ per real dimension)
is circularly symmetric complex Gaussian noise.
Under AWGN, $\mathbf{H}_t=\mathbf{I}$ for all $t$.
Under Rayleigh fading, the entries of $\mathbf{H}_t$ are i.i.d.\ $\mathcal{CN}(0,1)$ and we consider block fading
($\mathbf{H}_t=\mathbf{H}$ for all $t$). No explicit channel state information (CSI) or channel estimation is used. The receiver learns to decode directly from the received block.

To capture front-end nonidealities in the Rayleigh configuration, we extend the discrete\mbox{-}time complex baseband model with carrier\mbox{-}frequency offset (CFO) and phase noise via
\begin{equation}
\tilde{x}[t] = x[t]\,e^{j\theta[t]}, \qquad 
\theta[t] = 2\pi f_\Delta t + \phi[t],
\end{equation}
where \(f_\Delta = 0.01\) cycles per symbol and \(\phi[t]\) is a discrete-time Wiener process whose per-symbol innovation satisfies \(\Delta\phi[t]\sim\mathcal{N}(0,\sigma_\phi^2)\) with \(\sigma_\phi=0.005\) rad. I/Q imbalance is modeled as a linear distortion of the complex baseband signal for a \(+0.5\,\mathrm{dB}\) gain imbalance and a \(3^{\circ}\) quadrature phase skew. Converters are uniform mid\mbox{-}tread quantizers applied separately to I and Q with \(B_{\mathrm{TX}}=10\) and \(B_{\mathrm{RX}}=10\) effective bits, using step
\begin{equation}
\Delta(B)=\frac{2V_{\max}}{2^{B}-1}, \qquad B\in\{B_{\mathrm{TX}},B_{\mathrm{RX}}\},
\end{equation}
where $V_{\max}$ is set to $1$ in numerical results. Processing occurs in the following sequence: transmitter normalization, I/Q imbalance, carrier offset and phase noise, transmitter quantization, wireless channel, receiver quantization, neural receiver. AWGN-only baselines use an ideal front end without CFO, phase noise, I/Q imbalance, or quantization, whereas the Rayleigh results enable the impairment chain along with the fading MIMO matrices and additive noise.

The noise variance \(N_0\) is linked to the desired energy-per-bit to noise-density ratio $E_b/N_0$ through the effective rate
\begin{equation}
R_b = \frac{k}{n} \:\:\: \text{s.t.} \:\:\:\:
\frac{E_b}{N_0} = \frac{P}{R_b N_0},
\end{equation}
with \(P\) denoting the average energy per complex channel use induced by the normalization. For a specified $E_b/N_0$ in dB, the corresponding \(N_0\) follows from this relation. The mapping depends only on the information rate and not on \(N_t\) or \(N_r\), ensuring consistent bit energy across antenna configurations.

The receiver implements a decoder mapping
\begin{equation}
g_{\boldsymbol{\theta}}:\mathbb{C}^{N_r\times n} \rightarrow \mathbb{R}^k,\qquad
\boldsymbol{\ell} = g_{\boldsymbol{\theta}}(\mathbf{y}),
\end{equation}
with parameters \(\boldsymbol{\theta}\). The received sequence is first converted to an I/Q tensor of size \(n\times N_r\times 2\). To reduce sensitivity to large-scale gain fluctuations, a root-mean-square normalization is applied across time and receive antennas on a per-codeword basis. Further normalization along the flattened temporal dimension stabilizes feature magnitudes within each codeword. The normalized sequence is passed through a one-dimensional convolutional stem that lifts the I/Q input to a higher-dimensional feature space. Dilated residual temporal blocks then aggregate information across the full codeword and spatial dimensions, followed by squeeze-excitation style channel gating. Global average pooling produces a feature vector that a linear layer maps to \(k\) logits, with hard bits obtained by zero-thresholding. Increasing the number of receive antennas lengthens the input sequence, while the architecture and training remain unchanged for both SISO and MIMO.

\noindent\textbf{Encoder-decoder structure.}
The transmitter and receiver are implemented as residual temporal convolutional networks. At the transmitter, the bit vector is linearly projected to a length-$n$ feature map with $C_{\rm tx}$ channels and processed by a stack of residual 1-D convolutional blocks. A final $1\times1$ convolution produces $2N_t$ outputs corresponding to the I/Q components on each transmit antenna, followed by the energy normalization in (2). At the receiver, after the normalizations above, the length-$nN_r$ I/Q sequence passes through a convolutional stem and a deeper stack of residual temporal blocks with squeeze-excitation channel gating. Global average pooling and a linear layer produce the $k$ logits. The transmitter uses $C_{\rm tx}=64$ with $L_{\rm tx}=5$ residual blocks, and the receiver uses $C_{\rm rx}=128$ with $L_{\rm rx}=8$ residual blocks and kernel size $K=7$.

The transmitter and receiver are trained jointly by minimizing the expected bitwise binary cross-entropy between the logits and the ground-truth bits. For each codeword in a mini-batch, an $E_b/N_0$ value is drawn from a prescribed interval (e.g., \([-2,20]\) dB). The sampling distribution is biased towards intermediate SNRs where the BER curve is most sensitive, while still providing coverage at low and high SNRs. The corresponding noise variance is computed and applied to the channel realization. 

Both the encoder and decoder are implemented as convolutional neural networks of moderate depth and width, with a narrower stack at the transmitter and a deeper, wider stack at the receiver. The AE is trained on NVIDIA RTX PRO 6000 Blackwell Max-Q GPUs. The number of trainable parameters varies slightly with the number of antennas; e.g., when \(N_t = N_r = 4\), the encoder and decoder contain approximately \(0.30\) million and \(1.84\) million trainable parameters, respectively. Optimization uses a variant of Adam with weight decay, gradient clipping, and a learning-rate schedule that combines a short linear warmup with cosine decay to a reduced final step size. After training, BER is estimated by sweeping $E_b/N_0$ across the target range and simulating on the order of \(10^8\) bits per SNR point.

\begin{figure}[t!]
	\centering
	\includegraphics[width=0.8\columnwidth]{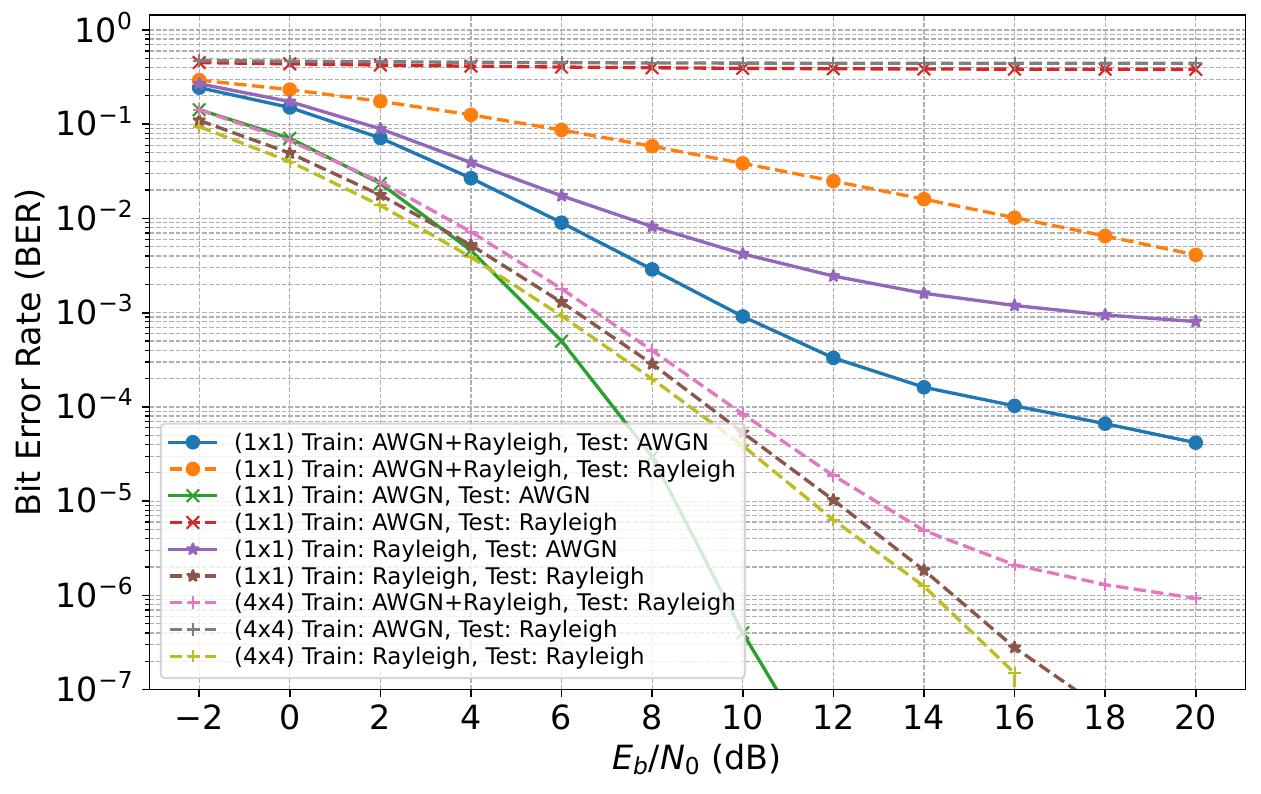}
	\caption{Effects of training-test environment mismatch on BER vs. $E_b/N_0$ for SISO and MIMO communications under AWGN and Rayleigh channels.}
	\label{fig:AESISOMIMO_mismatch}
\end{figure}

\begin{figure}[t!]
	\centering
	\includegraphics[width=0.8\columnwidth]{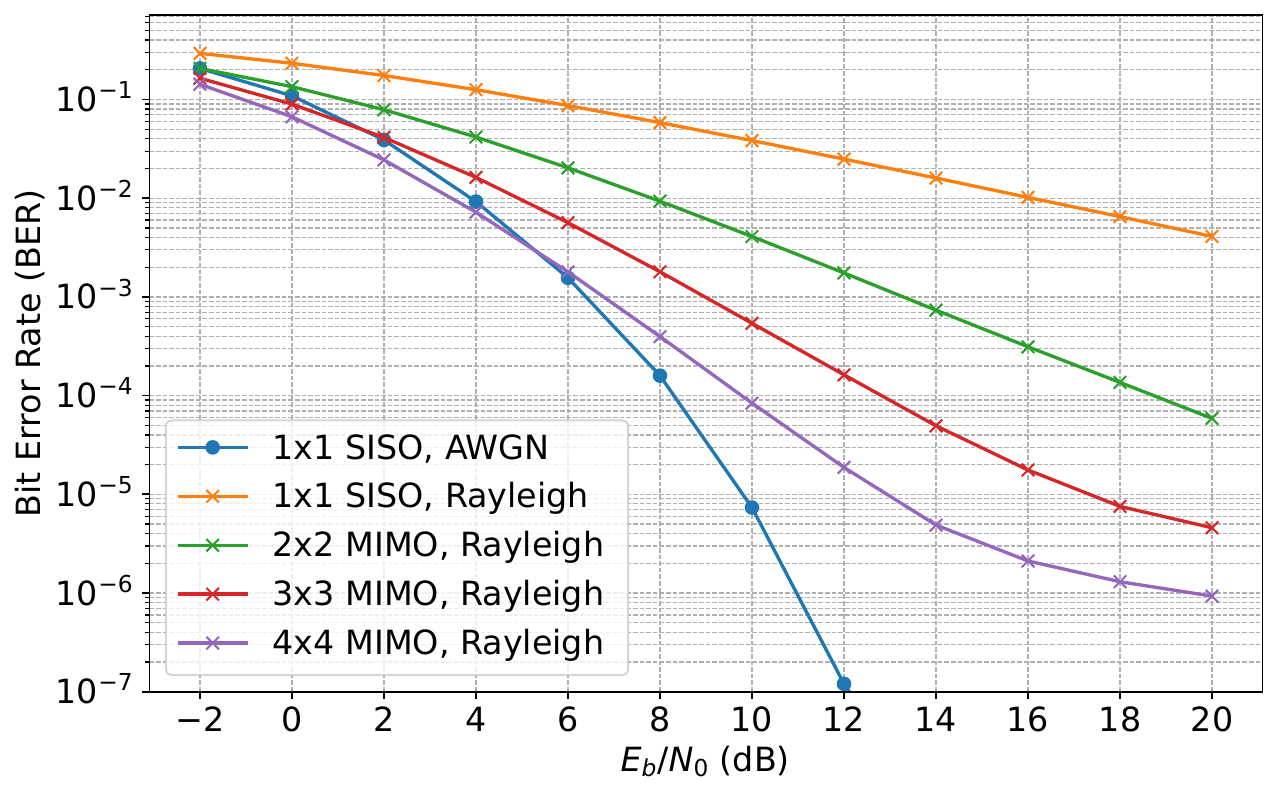}
	\caption{BER vs. $E_b/N_0$ for SISO and MIMO communications under AWGN and Rayleigh channels.}
	\label{fig:AESISOMIMO}
\end{figure}

\section{Channel-Driven Training-Test Mismatch}\label{sec:mismatch}

Real deployments in test time may not match the training-channel distribution. To study this, we train models on AWGN only, Rayleigh only (with impairments), and a mixed AWGN-Rayleigh curriculum, then evaluate each on both AWGN and Rayleigh (with impairments). Results are shown in Fig.~\ref{fig:AESISOMIMO_mismatch},  when $k = 8$ bits are mapped to $n = 16$ channel uses.

AWGN-only training does not transfer well to Rayleigh fading. Under fading, the BER remains high across the entire SNR range. Adding more antennas does not help much because the receiver was never trained to handle the amplitude and phase variations caused by fading. Rayleigh-only training, however, transfers well to AWGN. Under non-fading conditions, the BER decreases steadily as SNR increases, showing that features learned under the more challenging fading channel remain effective when the channel becomes simpler.

Mixed AWGN-Rayleigh training generalizes in both directions. On AWGN it closely tracks the AWGN-trained model, and on Rayleigh it maintains a steep BER decay with SNR, avoiding the saturation seen with AWGN-only training. With matched or mixed training, larger MIMO arrays further steepen the BER slopes and shift the rapid descent region to lower SNRs, whereas under mismatch additional antennas provide little benefit. In practice, matching test-time channel conditions during training is more important than increasing model size. AWGN-only training is insufficient for fading and hardware impairments, Rayleigh-plus-impairment-only training transfers conservatively to AWGN, and mixed training provides robust performance across both domains.

Continuing with mixed training, Fig.~\ref{fig:AESISOMIMO} shows BER for SISO over AWGN and Rayleigh channel with impairments, and for Rayleigh MIMO with \(N_t\!=\!N_r\!=\!2\), \(3\), and \(4\). The SISO AWGN curve exhibits a rapid BER reduction with increasing SNR, whereas the SISO Rayleigh curve decreases more gradually due to fading and front-end impairments. Increasing the number of antennas in the MIMO system shifts the BER curves toward lower SNRs and makes the transition from high to low BER sharper, demonstrating the benefits of spatial diversity. As the antenna count increases, the BER under Rayleigh fading with impairments moves closer to the AWGN performance, particularly at moderate and high SNRs, showing that the learned receiver and short-block coding strategy remain effective in more challenging channel conditions.

\begin{figure}[t!]
	\centering
	\includegraphics[width=0.8\columnwidth]{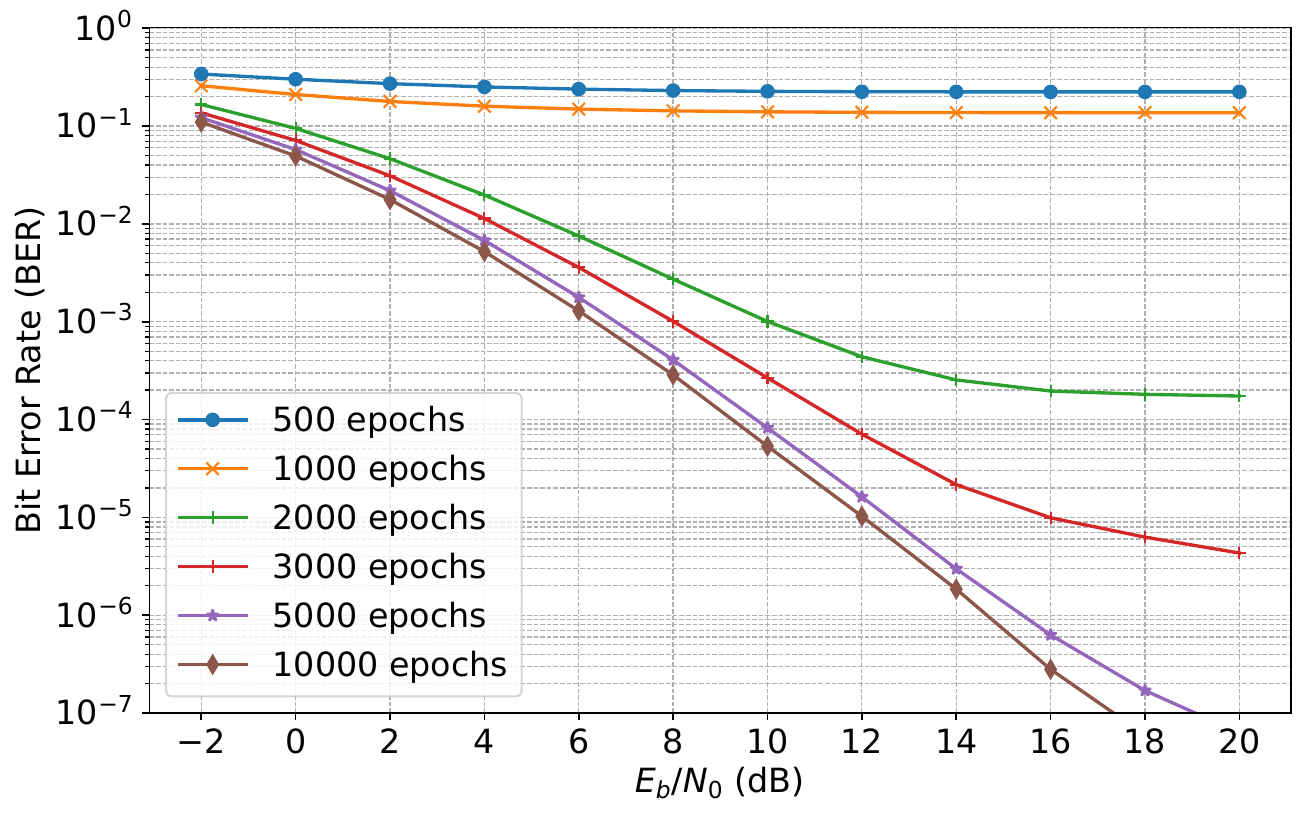}
	\caption{BER vs. $E_b/N_0$ for AE communications under Rayleigh channels with impairments as a function of the number of training epochs.}
\label{fig:AE4x4MIMORayleigh_effect_of_steps}
\vspace{-0.1cm}
\end{figure}

Fig.~\ref{fig:AE4x4MIMORayleigh_effect_of_steps} complements these results by isolating the training-time dimension for a fixed Rayleigh MIMO configuration. For a \(4\times 4\) Rayleigh channel with impairments, the BER curves after only a few training epochs remain far from their final performance. The BER reduction is gradual, occurs only at higher SNRs, and an error floor is still visible. As training continues, the curves shift toward lower SNRs and become steeper, eventually approaching the performance of the fully trained Rayleigh models in Fig.~\ref{fig:AESISOMIMO}. These results show that learning an effective end-to-end transmitter and receiver under Rayleigh fading with impairments requires significantly more training than under AWGN, with the largest performance improvements occurring at moderate SNRs.

Figs.~\ref{fig:AESISOMIMO_mismatch}-\ref{fig:AE4x4MIMORayleigh_effect_of_steps} reveal two issues motivating transfer learning. First, naive deployment of an AWGN-trained system under Rayleigh fading with impairments fails due to channel mismatch, even when source-domain performance is strong. Second, closing this gap through from-scratch training requires long training schedules and substantial computation. We therefore use the AWGN-trained model as a strong initialization and perform a short adaptation phase tailored to Rayleigh fading with impairments, correcting the mismatch at a fraction of the cost of full retraining.

\section{Transfer Learning for Spectrum Adaptation} \label{sec:lowlatency}

The encoder-decoder pair above can be trained for any channel, but learning a robust transmitter and receiver from random initialization in MIMO Rayleigh fading with impairments is significantly more demanding than under AWGN. In this setting, the system must jointly discover suitable signaling patterns and an effective equalization behavior for random complex channel matrices. To reduce training time and computational cost, we adopt transfer learning: the transmitter and receiver are first trained jointly under AWGN, and then only a subset of layers is adapted to Rayleigh fading with impairments while the rest remain fixed. Let \(\boldsymbol{\psi} = (\boldsymbol{\theta},\boldsymbol{\phi})\) denote the complete set of transmitter and receiver parameters.

Under AWGN, the channel follows (\ref{eq:channel}) with $\mathbf{H}_t=\mathbf{I}$, so the conditional distribution of the received sequence is
\begin{equation}
p_{\mathrm{AWGN}}(\mathbf{y}\mid\mathbf{x})=\mathcal{CN}(\mathbf{y};\mathbf{x},N_0\mathbf{I}),
\end{equation}
where $\mathbf{y}$ denotes the stacked received vector over the codeword. When training over a distribution of SNR values, the objective is to minimize the SNR-averaged risk
\begin{equation}
\min_{\boldsymbol{\psi}} \;
\mathbb{E}_{\mathbf{b},\gamma,\mathbf{w}}\Big[
\ell\big(\mathbf{b}, g_{\boldsymbol{\theta}}(f_{\boldsymbol{\phi}}(\mathbf{b}) + \mathbf{w}(\gamma))\big)
\Big],
\end{equation}
where \(\ell(\cdot,\cdot)\) is the bitwise cross-entropy loss, \(\gamma\) indexes the sampled $E_b/N_0$, and $\mathbf{w}(\gamma)$ denotes AWGN with variance determined by the effective rate and the SNR value $\gamma$. This AWGN-only training yields parameters \(\boldsymbol{\psi}_\text{AWGN}\) that form a well-separated, power-normalized constellation and a receiver that robustly denoises and classifies over a wide SNR range.

Under Rayleigh fading, the channel follows (\ref{eq:channel}) with 
$\mathbf{H}_t=\mathbf{H}$ constant over the codeword (block fading), 
where $\mathbf{H}\in\mathbb{C}^{N_r\times N_t}$ has i.i.d.\ 
$\mathcal{CN}(0,1)$ entries. The resulting conditional density can be written as
\begin{equation}
p_{\mathrm{Rayleigh}}(\mathbf{y}\mid\mathbf{x})
= \mathbb{E}_{\mathbf{H}}\Big[
\mathcal{CN}\big(\mathbf{y}; \mathcal{H}(\mathbf{x},\mathbf{H}), N_0\mathbf{I}\big)
\Big],
\end{equation}
where $\mathcal{H}(\mathbf{x},\mathbf{H})$ denotes the noiseless MIMO output produced by the channel matrix $\mathbf{H}$. In the impaired Rayleigh configuration used in our experiments, the front-end distortion and quantization chain of Sec.~\ref{sec:system_model} is applied to the transmitted and received samples in addition to this fading channel and AWGN (for notational simplicity this is absorbed into the channel type). Compared to AWGN, Rayleigh fading introduces random rotations, scalings, and cross-antenna couplings in the received feature space. A receiver trained only on AWGN effectively approximates the Bayes decision rule for an identity channel and tends to misclassify signals when evaluated under random \(\mathbf{H}\), often exhibiting a BER floor that persists even at moderate and high $E_b/N_0$. Training a fresh transmitter-receiver pair directly under Rayleigh fading would solve
\begin{equation}
\min_{\boldsymbol{\psi}} \;
\mathbb{E}_{\mathbf{b},\gamma,\mathbf{H},\mathbf{w}}\Big[
\ell\big(\mathbf{b},
g_{\boldsymbol{\theta}}(\mathcal{H}(f_{\boldsymbol{\phi}}(\mathbf{b}),\mathbf{H}) + \mathbf{w}(\gamma))\big)
\Big],
\end{equation}
starting from random initialization. This requires exploring the full parameter space and jointly learning both the signaling structure and an implicit equalization behavior for random matrix-valued channels, which is computationally expensive.

To avoid this, we treat \(\boldsymbol{\psi}_\text{AWGN}\) as a prior and adapt only a small subset of layers to the Rayleigh setting. The transmitter is viewed as a sequence of blocks with parameters \(\boldsymbol{\phi}^{(1)},\dots,\boldsymbol{\phi}^{(L_{\text{enc}})}\), ordered from the bit input to the waveform output, and the receiver as a sequence of blocks with parameters \(\boldsymbol{\theta}^{(1)},\dots,\boldsymbol{\theta}^{(L_{\text{dec}})}\), ordered from the channel observations to the logits. We select integers \(M_{\text{enc}}\) and \(M_{\text{dec}}\) and declare that only the first \(M_{\text{enc}}\) transmitter blocks and the first \(M_{\text{dec}}\) receiver blocks will be updated during adaptation, while all deeper blocks remain fixed. Formally,
\begin{IEEEeqnarray}{l}
\hspace{-0.5cm} \boldsymbol{\psi}_{\text{adapt}}
= \big(\boldsymbol{\phi}^{(1)},\dots,\boldsymbol{\phi}^{(M_{\text{enc}})},
\boldsymbol{\theta}^{(1)},\dots,\boldsymbol{\theta}^{(M_{\text{dec}})}\big),
\\
\hspace{-0.5cm} \boldsymbol{\psi}_{\text{frozen}}
= \big(\boldsymbol{\phi}^{(M_{\text{enc}}+1)},\dots,\boldsymbol{\phi}^{(L_{\text{enc}})},
\boldsymbol{\theta}^{(M_{\text{dec}}+1)},\dots,\boldsymbol{\theta}^{(L_{\text{dec}})}\big).
\end{IEEEeqnarray}
The spectrum adaptation problem then becomes
\begin{equation}
\min_{\boldsymbol{\psi}_\text{adapt}} \;
\mathbb{E}_{\mathbf{b},\gamma,\mathbf{H},\mathbf{w}}\Big[
\ell\big(\mathbf{b},
g_{\boldsymbol{\theta}}(\mathcal{H}(f_{\boldsymbol{\phi}}(\mathbf{b}),\mathbf{H}) + \mathbf{w}(\gamma))\big)
\Big],
\end{equation}
subject to the constraint \(\boldsymbol{\psi}_\text{frozen} = \boldsymbol{\psi}_\text{AWGN,frozen}\), and with initialization
\(\boldsymbol{\psi}_\text{adapt} = \boldsymbol{\psi}_\text{AWGN,adapt}\). Here, $\boldsymbol{\phi}$ and $\boldsymbol{\theta}$ contain both the adapted and frozen parameter subsets, with
$\boldsymbol{\psi}=(\boldsymbol{\theta},\boldsymbol{\phi})
=(\boldsymbol{\psi}_{\mathrm{adapt}},\boldsymbol{\psi}_{\mathrm{frozen}})$. Deeper transmitter and receiver layers are kept at their AWGN-trained values, while only the selected early blocks are updated. On the transmitter side, these early blocks learn fading- and impairment-aware preprocessing of the bit patterns, generating internal features better matched to the target-channel statistics before the frozen backbone maps them to a codeword. On the receiver side, the early blocks learn to partially invert the random linear mixing and denoise the observations, producing an AWGN-like representation for the unchanged deeper blocks, which retain their learned bit-decision function. Thus, most of the AWGN-learned structure is preserved, and only the selected early blocks adapt to the new channel conditions.

Let \(P_{\text{tot}}\) denote the total parameter count and \(P_{\text{adapt}}\) the subset in \(\boldsymbol{\psi}_{\text{adapt}}\). If from-scratch training requires \(T_{\text{scratch}}\) steps to reach a target error rate under Rayleigh fading with impairments, the update cost scales as \(T_{\text{scratch}}P_{\text{tot}}\). With transfer learning, AWGN pretraining incurs a one-time cost \(T_{\text{AWGN}}P_{\text{tot}}\), while adaptation costs only \(T_{\text{adapt}}P_{\text{adapt}}\), both far smaller than full retraining. Thus, adaptation to Rayleigh channel with impairments achieves low BER at a small fraction of the full cost. The AWGN-trained AE serves as an SNR-robust baseline, and the proposed layer-wise adaptation provides an efficient path to fading-robust operation.

\begin{figure}[t!]
	\centering
	\includegraphics[width=0.8\columnwidth]{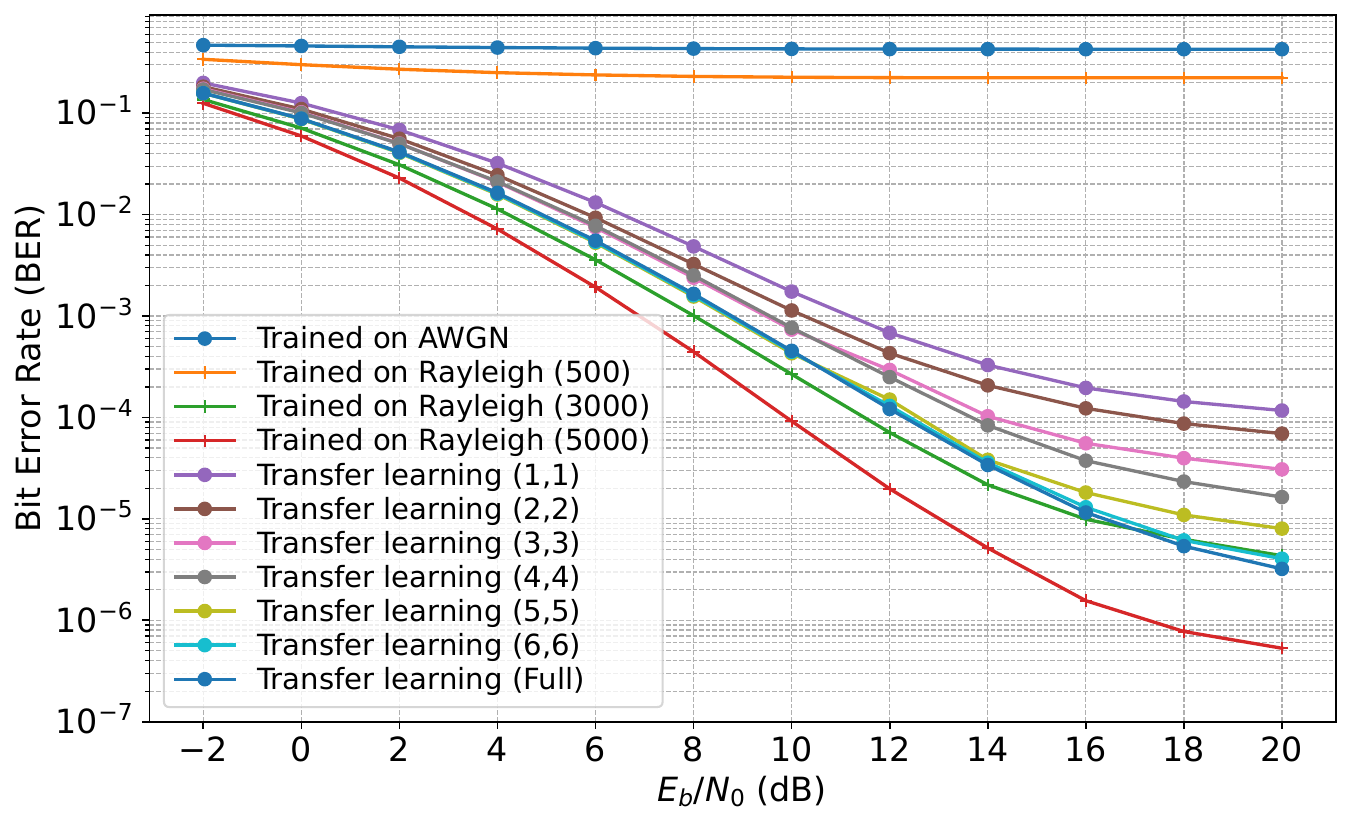}
	\caption{BER vs. $E_b/N_0$ for transfer learning.}
	\label{fig:TL_BER_AWGN_to_4x4Ray_BER_k8}
\end{figure}

Fig.~\ref{fig:TL_BER_AWGN_to_4x4Ray_BER_k8} shows BER versus $E_b/N_0$ for a \(4{\times}4\) Rayleigh channel with impairments when starting from an AWGN-trained transmitter-receiver and adapting different layer subsets. In Fig.~\ref{fig:TL_BER_AWGN_to_4x4Ray_BER_k8}, \((n_e,n_d)\) denotes updating only the first \(n_e\) encoder blocks and the first \(n_d\) decoder blocks during fine-tuning, with all remaining layers frozen at their AWGN-pretrained values. The AWGN-only model (no adaptation) performs poorly under Rayleigh channel with impairments, maintaining high BER across SNR. As \((n_e,n_d)\) increases from \((1,1)\), the BER curves move monotonically toward that of full adaptation. BER decreases more rapidly with SNR, the curves shift to lower SNRs, and the high-SNR error floor is progressively removed. With a moderate number of updated blocks, performance nearly matches the fully adapted model, indicating that updating only a fraction of layers suffices to significantly reduce BER.

Unlike transfer learning in computer vision and related applications, where the labels change but the input statistics remain broadly similar, our setting keeps the bit semantics fixed while the channel characteristics vary. Empirically, we find that adapting the early layers, which act as a learned equalizer or front end, provides larger performance gains than tuning the final layers, suggesting that effective channel adaptation primarily requires retuning the parts of the network that operate closest to the received signal.

\begin{figure}[t!]
  \centering

  \begin{subfigure}{0.49\columnwidth}
    \centering
    \includegraphics[width=\linewidth]{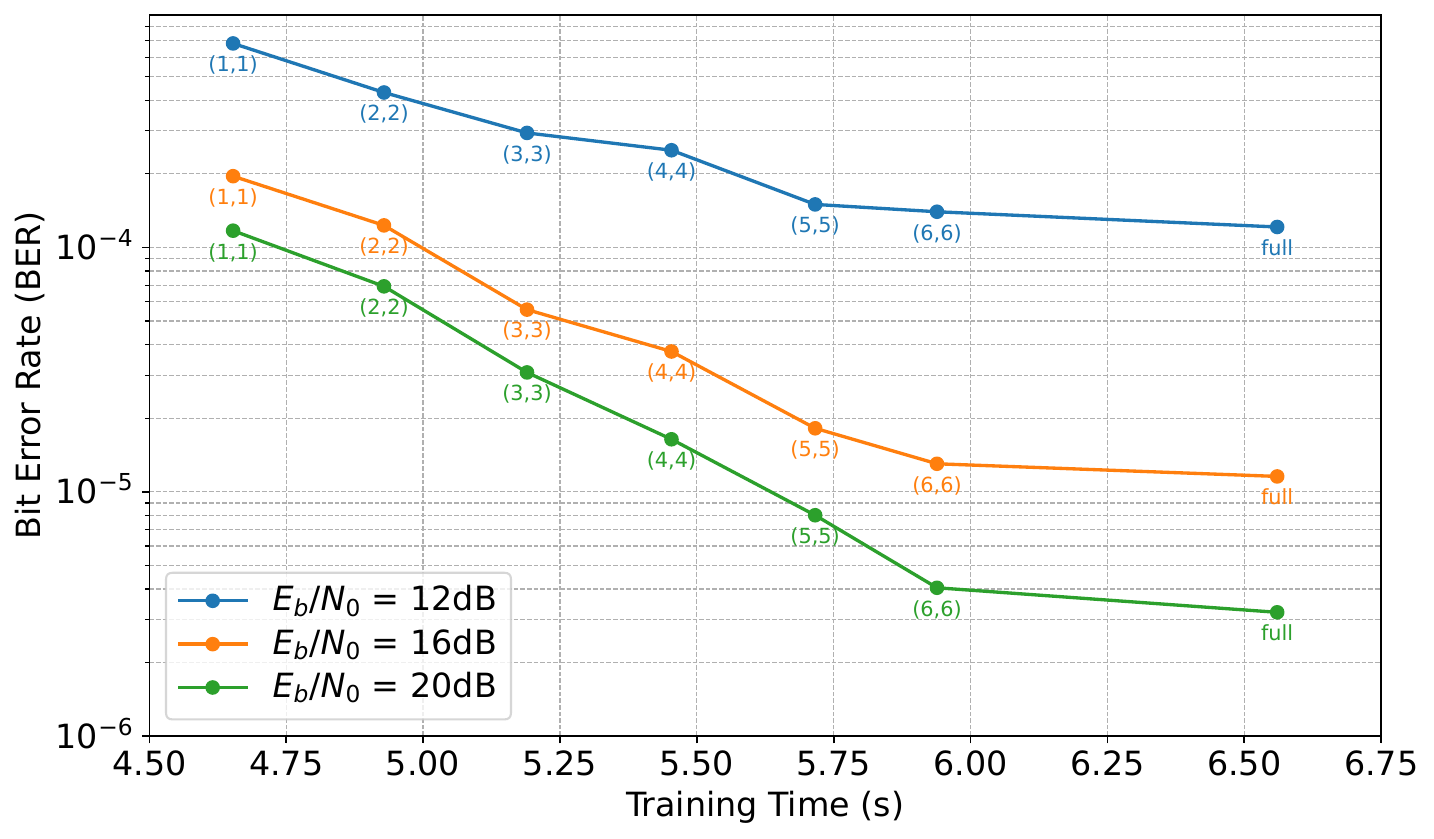}
    \caption{BER vs. training time.}
    \label{fig:TL_time}
  \end{subfigure}\hfill
  \begin{subfigure}{0.49\columnwidth}
    \centering
    \includegraphics[width=\linewidth]{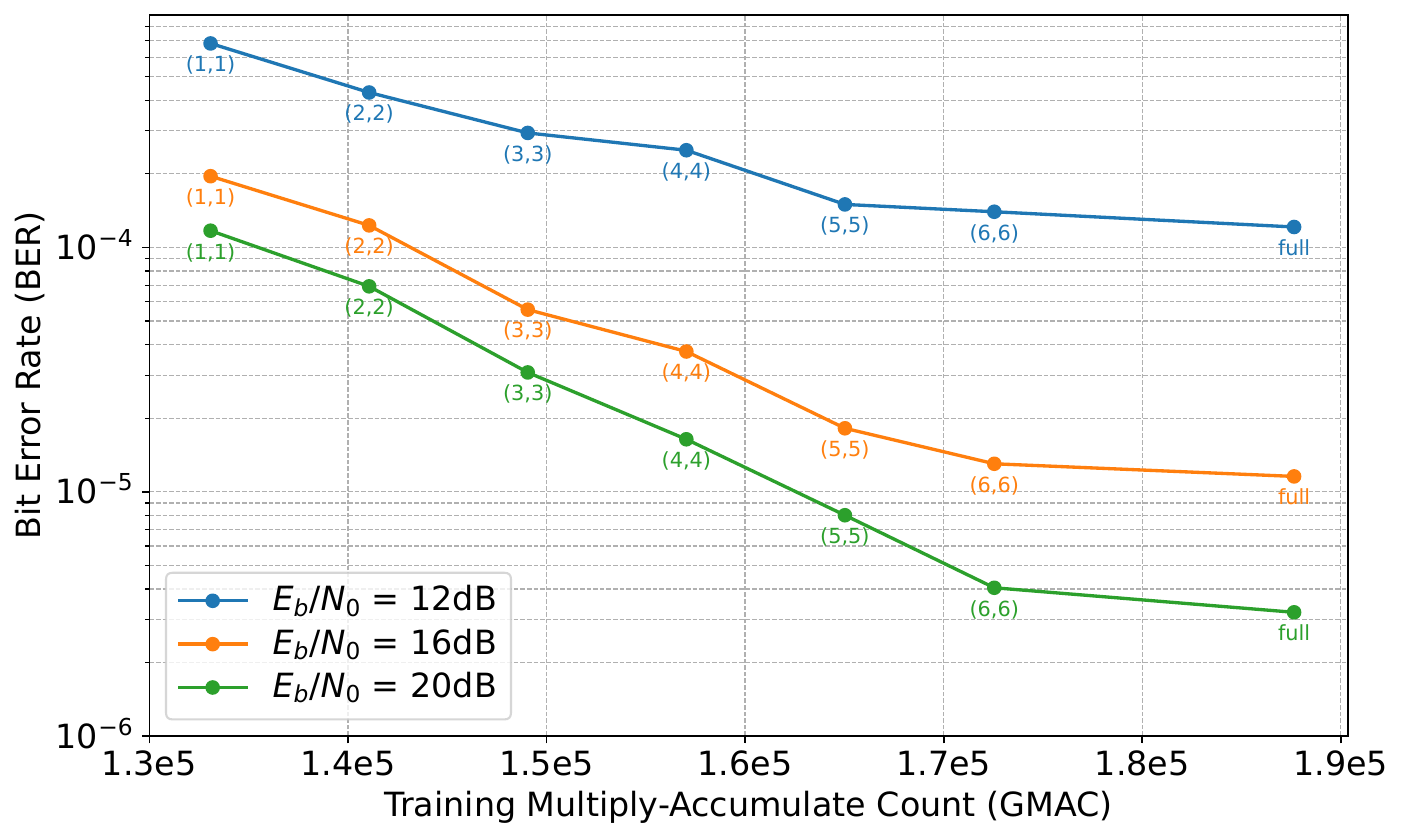}
    \caption{BER vs. training MAC.}
    \label{fig:TL_MAC}
  \end{subfigure}

  \vspace{0.6em}

  \begin{subfigure}{0.49\columnwidth}
    \centering
    \includegraphics[width=\linewidth]{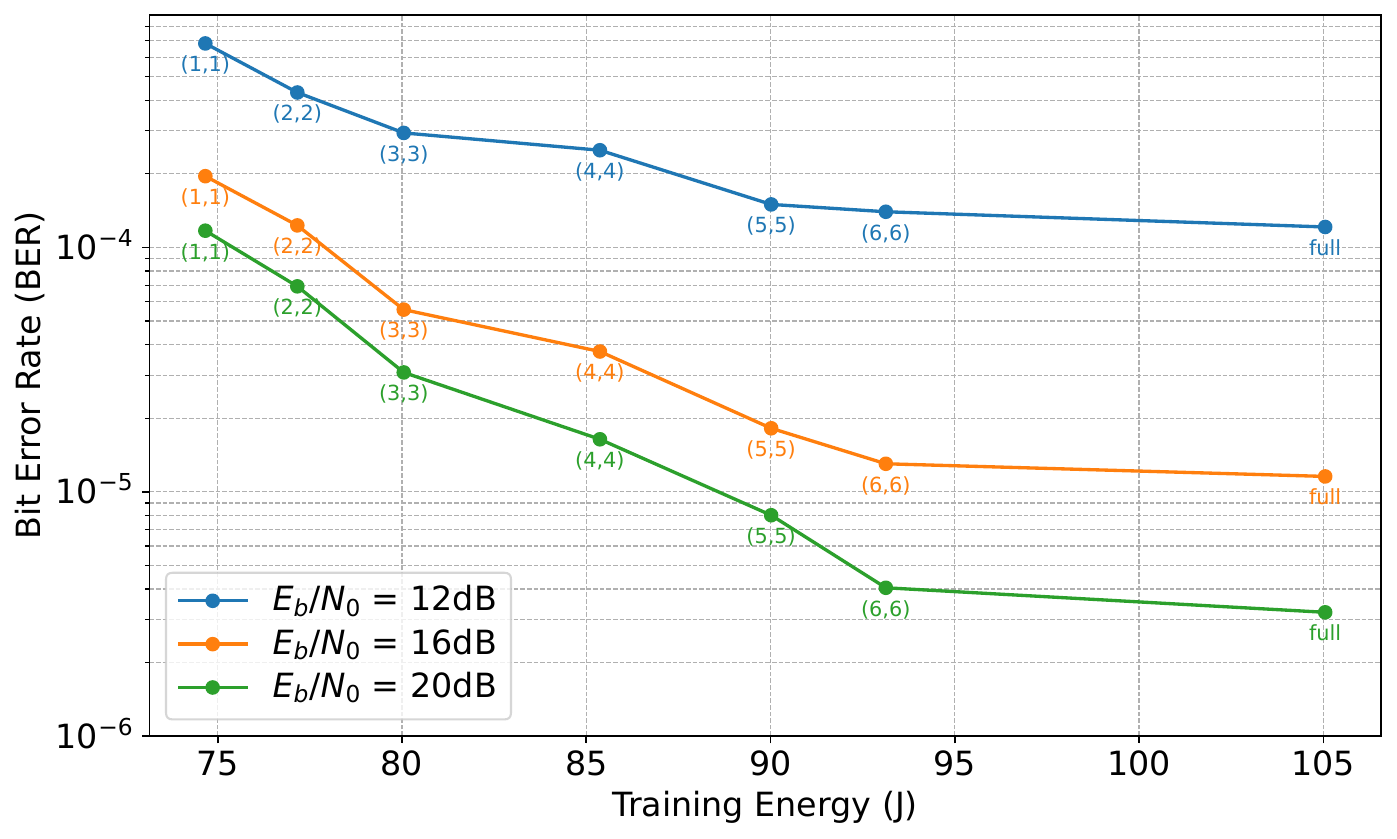}
    \caption{BER vs. training energy.}
    \label{fig:TL_energy}
  \end{subfigure}\hfill
  \begin{subfigure}{0.49\columnwidth}
    \centering
    \includegraphics[width=\linewidth]{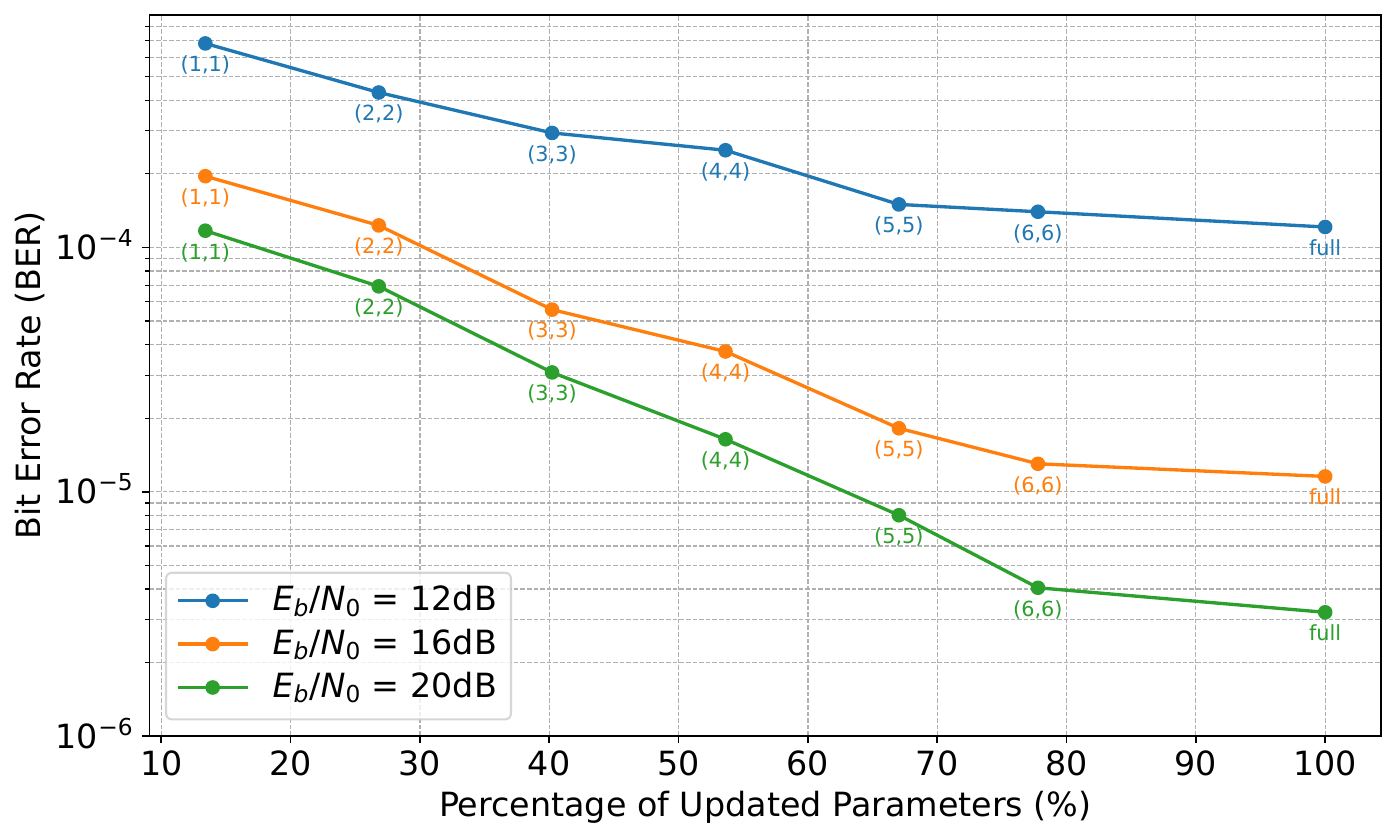}
    \caption{BER vs. training parameters.}
    \label{fig:TL_par}
  \end{subfigure}

  \caption{Transfer learning results: (a) BER vs. training time; (b) BER vs. training MAC; (c) BER vs. training energy; (d) BER vs. training parameters.}
  \label{fig:TL_2x2}
\end{figure}

Figs.~\ref{fig:TL_time}-\ref{fig:TL_par} summarize the resource-performance trade-offs when adapting a \(4{\times}4\) Rayleigh system with impairments from an AWGN initialization for different \((n_e,n_d)\). Fig.~\ref{fig:TL_time} plots BER at $E_b/N_0 \in\{12,16,20\}$~dB versus training time. Updating more blocks increases time while monotonically reducing BER, with intermediate \((n_e,n_d)\) already near the fully updated case. Fig.~\ref{fig:TL_MAC} reports the same trend versus total MAC count, confirming MACs as a reliable proxy for adaptation time. Fig.~\ref{fig:TL_energy} reports BER versus training energy (measured on GPU with PyNVML) and exhibits a similar pattern, where moderate adaptation cost achieves low BER unattainable by from-scratch training under the same step budget. Fig.~\ref{fig:TL_par} reports BER versus the fraction of updated parameters, showing that low BER is achievable by adapting only a subset of layers rather than retraining the full transmitter-receiver.

This procedure yields lower BER than fully retrained models after relatively few adaptation steps, while requiring far fewer gradient updates and less time and energy. The AWGN-trained AE therefore serves as a generic, SNR-robust base communications system, and transfer learning provides an efficient route to fading-robust performance in more complex channel conditions. To further control adaptation cost, we employ a budget-constrained parameter selection scheme that updates only a subset of transmitter and receiver parameters under a single global budget. The encoder and decoder are partitioned into disjoint parameter groups, typically aligned with convolutional or residual blocks. Let \(\{\mathcal{P}_u\}_{u\in\mathcal{G}}\) denote these groups, where each \(\mathcal{P}_u\) contains \(|\mathcal{P}_u|\) scalar parameters.

Starting from the AWGN-pretrained model, we perform a short probe phase under the Rayleigh channel with impairments. During this phase, the model is exposed to a small number of mini-batches, and for each group \(\mathcal{P}_u\) we accumulate the squared \(\ell_2\)-norm of its gradient, i.e.,
\begin{equation}
I_u \;\triangleq\; \sum_{t} \big\|\nabla_{\mathcal{P}_u} \ell_t\big\|_2^2,
\end{equation}
where \(\ell_t\) is the loss on mini-batch \(t\). The term \(I_u\) serves as an importance score. Groups with larger values are expected to yield greater loss reduction when made trainable during adaptation.
The total number of scalar parameters in the joint transmitter-receiver model is
\begin{equation}
P_{\mathrm{tot}} \;=\; \sum_{u\in\mathcal{G}} |\mathcal{P}_u|.
\end{equation} 
For a prescribed fraction \(\alpha\in(0,1]\), we define a global budget
\begin{equation}
B \;=\; \alpha\,P_{\mathrm{tot}}.
\end{equation}
The groups \(\{\mathcal{P}_u\}\) are sorted by decreasing importance score \(I_u\) and marked trainable until the cumulative parameter count of the selected set \(\mathcal{S}\subset\mathcal{G}\) first exceeds \(B\), i.e.,
\begin{equation}
\sum_{u\in\mathcal{S}} |\mathcal{P}_u| \le B,
\qquad
\sum_{u\in\mathcal{S}\cup\{v\}} |\mathcal{P}_u| > B
\;\text{for any } v\notin\mathcal{S}.
\end{equation}

All parameters in groups $u\in\mathcal{S}$ are updated during Stage~2 training on Rayleigh fading with impairments, while the remaining parameters are frozen at their AWGN-pretrained values. This global-budget selection allocates adaptation capacity between encoder and decoder based on their measured probe-phase importance rather than enforcing a fixed split. Choosing $\alpha$ on the order of a few tens of percent updates only a modest fraction of parameters while closing most of the gap to fully retrained Rayleigh-plus-impairment models. This approach achieves low BERs with substantially reduced adaptation time, computation, and energy cost.

\begin{figure}[t!]
	\centering
	\includegraphics[width=0.8\columnwidth]{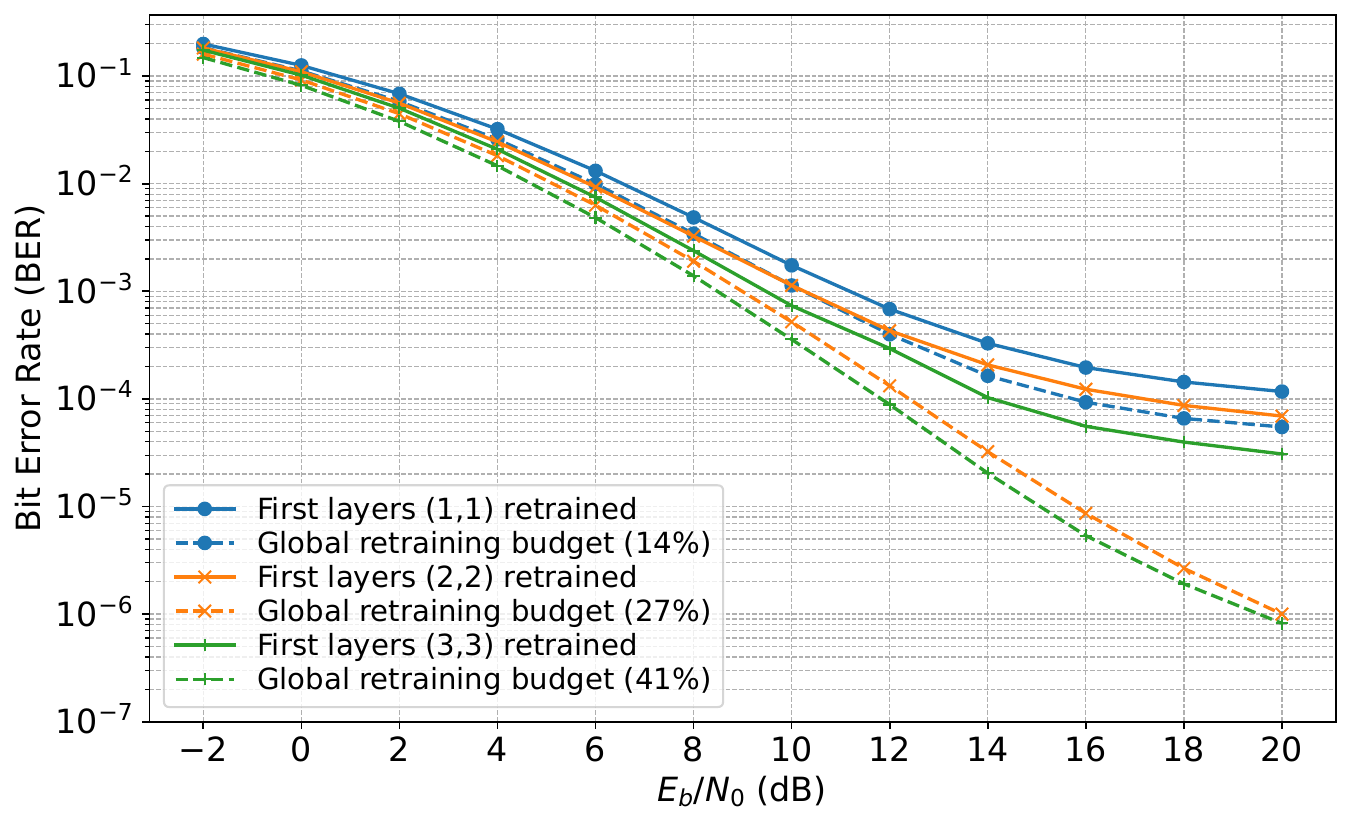}
	\caption{BER vs. $E_b/N_0$ when either (i) first layers of encoder and decoder are retrained, or (ii) retrained layers are selected subject to a retraining budget.}
	\label{fig:Global_budget_k8_new_2}
\end{figure}

Beyond the simple first-\((n_e,n_d)\) schemes, we compare against the global-budget selection strategy by adjusting the retrained parameter fraction. Results are shown in Fig.~\ref{fig:Global_budget_k8_new_2}. Updating the first \((1,1)\), \((2,2)\), and \((3,3)\) blocks retrains about \(13.41\%\), \(26.82\%\), and \(40.22\%\) of parameters. With global budgets of \(14\%\), \(27\%\), and \(41\%\), the data-driven policy instead uses \(13.58\%\), \(24.38\%\), and \(40.59\%\), respectively, and consistently improves BER across $E_b/N_0$. For the \(14\%\) budget, the selected set includes encoder groups \(\{4,5\}\) and decoder groups \(\{0,8\}\); for \(27\%\), the encoder again uses \(\{4,5\}\) while the decoder expands to \(\{0,6,8\}\); and for \(41\%\), the policy activates encoder groups \(\{2,3,4,5\}\) and decoder groups \(\{0,3,6,8\}\). The global-budget policy departs from the first-layers-only rule, reallocating a similar (or smaller) parameter budget across early and deeper blocks, particularly in the receiver, and achieving lower BER than prefix-based schemes.

\section{Conclusion} \label{sec:conclusion}
This paper studied end-to-end learned transmitter–receiver designs and their sensitivity to channel mismatch, showing that models trained only on AWGN channels degrade markedly when deployed over Rayleigh fading with impairments. To address the cost of full retraining, we proposed a transfer-learning approach that initializes from an AWGN-pretrained system and fine-tunes only a subset of transmitter and receiver layers while freezing the remainder. This partial adaptation removes the mismatch-induced error and achieves low BER with far fewer training steps, computation, energy, and parameter updates than full retraining. We also introduced a budget-constrained layer-selection procedure that identifies the most informative parameters to update under a global retraining budget, further improving BER under comparable adaptation cost. The results suggest a practical guideline for learned physical layers: preserve the signal geometry and deep decision structure learned under AWGN, and adapt lightweight front-end processing to account for fading with impairments.

\bibliographystyle{IEEEtran}

\bibliography{references_tl_new}

\end{document}